\documentclass[reprint,twocolumn,superscriptaddress,secnumarabic,amssymb,nobibnotes,aps,prb]{revtex4-2}
\usepackage{extarrows}
\usepackage{amsmath}
\usepackage{graphicx}
\usepackage{verbatim}
\usepackage{color}
\usepackage{subfigure}
\usepackage{hyperref}
\usepackage{longtable}
\usepackage{textcomp}
\usepackage{hhline}
\DeclareSymbolFont{symbolsC}{U}{txsyc}{m}{n}
\SetSymbolFont{symbolsC}{bold}{U}{txsyc}{bx}{n}
\DeclareFontSubstitution{U}{txsyc}{m}{n}
\DeclareMathSymbol{\multimapboth}{\mathrel}{symbolsC}{"13}
\DeclareMathSymbol{\multimapdotboth}{\mathrel}{symbolsC}{"16}
\DeclareMathSymbol{\multimapdotbothA}{\mathrel}{symbolsC}{"17}
\DeclareMathSymbol{\multimapdotbothB}{\mathrel}{symbolsC}{"18}
\DeclareMathOperator{\sgn}{sgn}
\DeclareMathOperator{\ssin}{\smash{\sin}}

\begin{document}

\title{Semiclassical model of magnons in double-layered antiferromagnets}

\author{Seo-Jin Kim}                  
\affiliation{Max Planck Institute for Chemical Physics of Solids, D-01187 Dresden, Germany}
\author{Zden\v{e}k Jir\'{a}k}         
\affiliation{Institute of Physics, Czech Academy of Sciences, Cukrovarnick\'{a} 10, 162 00 Praha 6, Czechia}
\author{Ji\v{r}\'{i} Hejtm\'{a}nek}   
\affiliation{Institute of Physics, Czech Academy of Sciences, Cukrovarnick\'{a} 10, 162 00 Praha 6, Czechia}
\author{Karel Kn\'{i}\v{z}ek}         
\affiliation{Institute of Physics, Czech Academy of Sciences, Cukrovarnick\'{a} 10, 162 00 Praha 6, Czechia}
\author{Helge Rosner}                 
\affiliation{Max Planck Institute for Chemical Physics of Solids, D-01187 Dresden, Germany}
\author{Kyo-Hoon Ahn}                 
\email[Contact author: ]{kyohoon.ahn@fzu.cz}
\affiliation{Institute of Physics, Czech Academy of Sciences, Cukrovarnick\'{a} 10, 162 00 Praha 6, Czechia}

\begin{abstract}
The stability and magnonic properties of double-layered antiferromagnets are investigated using two model systems, a linear chain (LC) and a more complex chain of railroad trestle (RT) geometry, and the results are confronted with properties of the real material CrN.
The spin-paired order ($\cdots{+}{+}{-}{-}\cdots$) in LC requires alternating ferromagnetic and antiferromagnetic (AFM) exchanges, whereas in RT, an analogous order remains stable even when all interactions are AFM within certain analytical constraints.
The rock-salt structure of CrN evokes clear magnetic frustration since Cr atoms in a face-centered cubic lattice form links to twelve nearest neighbors (NNs) all equivalent and AFM.
Nonetheless, the magnetostructural transition to an orthorhombically distorted phase below $T_\text{N} = 287$~K leads to four different NN Cr--Cr distances and consequently, to a large diversification of the exchange strength, which suppresses the frustration and allows for stable double-layered AFM order of CrN.
This behavior originates from a competition at each NN link between Cr--Cr direct exchange and 90\textdegree{} Cr--N--Cr superexchange, both exhibiting specific power-law dependences on the interatomic distance.
Finally, based on the \textit{ab initio} calculated exchange parameters, the magnon spectrum and temperature evolution of ordered magnetic moments are derived.
\end{abstract}

\maketitle

\section{Introduction}
Antiferromagnetism is generally associated with the formation of two sublattices in a solid whose magnetic moments have equal magnitudes but in opposite orientations.
There are always exchange interactions between pairs of atomic spins, characterized by a negative (antiferromagnetic; AFM) exchange parameter $J$, that are responsible for the inter-sublattice coupling, while it is commonly believed that the intra-sublattice interactions are either insignificant or are characterized by a positive (ferromagnetic; FM) $J$.
The first eventuality is typical for AFM ordering in model lattices of linear chain (LC), planar square, simple cubic, and body-centered cubic types.
However, the latter requirement of positive $J$ is only true for a single pair of spins, while it is uncertain for complicated spin arrangements in condensed matter \cite{Zener1953}.
It would be of great help in understanding spin systems if a clear illustration of some non-intuitive behavior were presented.

We propose double-layered antiferromagnets as a case that goes beyond the conventional perspective by showing the AFM exchange to exist even inside the FM sublattices.
To provide a realistic example, we consider AFM CrN with an orthorhombically distorted rock-salt structure below the N\'{e}el temperature of $T_\text{N} = 287$~K \cite{Wang2023}.
Its spin arrangement is formed by positively and negatively oriented FM layers in the crystallographic $bc$ plane that alternate in the desired paired sequence of $|{+}{+}{-}{-}\rangle \equiv \cdots{+}{+}{-}{-}\cdots$ along the $a$ direction, with ordered local Cr moments making 2.35~$\mu_\text{B}$ as determined by neutron diffraction \cite{Corliss1960,Gui2022}.
Electronic structure calculations within the local density approximation (LDA) identify CrN as intrinsically metallic; however, applying an on-site Coulomb repulsion $U > 0.5$~eV opens a band gap (see Supplemental Material of Ref.~\cite{Gui2022} for details).
For $U = 4.1$~eV employed in this work, the gap makes $\sim$0.5~eV and the magnetic moment inside the Cr spheres has been calculated to be 2.85~$\mu_\text{B}$, suggesting an atomic spin close to the ideal value $S = 3/2$.

\begin{figure}[b]
\includegraphics[scale=0.34]{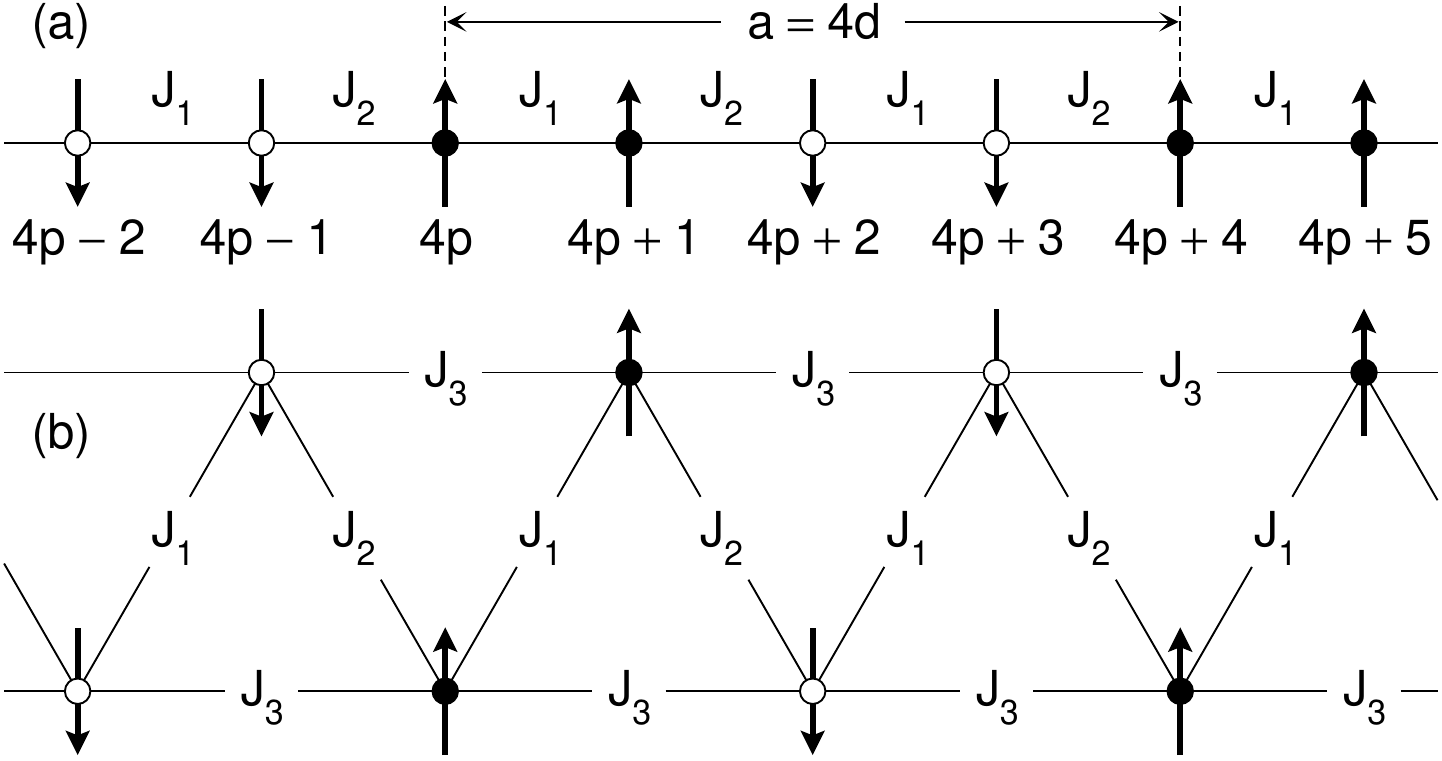}
\caption{
(a) One-dimensional (1D) model of double-layered antiferromagnets in the linear chain structure.
The spin arrangement is represented by the arrows (circles) which are upwards ($\bullet$) for $S^z = +S$ and downwards ($\circ$) for $-S$.
The exchange parameters $J_1$ and $J_2$ are for the intra- ($\multimapdotboth$,~$\multimapboth$) and inter-sublattice ($\multimapdotbothB$) interactions, respectively.
All sites have the same spacing $d$ and are indexed by $4p + n$, where $n = 0$, 1, 2, and 3 in the $p$th unit cell.
(b) 1D model of the railroad trestle type, possessing an additional exchange parameter $J_3$.
Each site has one nearest neighbor (NN) of same spin (related by $J_1$) and three NNs of opposite spin (related by $J_2$ and $J_3$).
}
\label{fig:struct}
\end{figure}

To elucidate the complex interplay among different exchange interactions in double-layered antiferromagnets, the present study proceeds as follows.
First, the equations of spin motion for two one-dimensional (1D) models of double-layered AFM order are solved and the corresponding magnon dispersion relations are determined analytically.
The model shown in Fig.~\ref{fig:struct}(a) represents a simple LC.
Our treatment essentially follows what is standardly presented for ferro- and antiferro-magnons in textbooks, see e.g., Ref.~\cite{Kittel2004}.
The spin-paired arrangement is described using a four-atom unit cell with periodic boundary conditions, assuming a N\'{e}el type ground state in which the $|{+}{+}{-}{-}\rangle$ spin configuration with $S^z = +S$ and $-S$ is stabilized by an alternation of two exchange parameters, the FM ($J_1 > 0$) and AFM ($J_2 < 0$) ones.
The model in Fig.~\ref{fig:struct}(b) represents a more complex 1D system that can possibly be named as railroad trestle (RT).
Such a lattice better reflects the real situation in three-dimensional (3D) AFM CrN and includes now three exchange parameters, which can be under some conditions all negative and of comparable strength, still keeping the $|{+}{+}{-}{-}\rangle$ ground state stable.
In the next step, the exchange parameters of CrN are evaluated from \textit{ab initio} calculations for three variants of the crystal structure, from a hypothetical strictly cubic rock-salt type to a fully optimized orthorhombic one.
The structural distortion leads to a diversification of nearest-neighbor (NN) Cr--Cr distances and to the emergence of four AFM exchange parameters whose values correlate strongly with these distances.
The origin of such correlation is presumably related to the interplay of AFM Cr--Cr direct exchange and FM 90\textdegree{} Cr--N--Cr superexchange.

\section{Theoretical background}
We introduce the time evolution of a spin under the action of torque forces, which is required to solve the magnon spectrum.
Consider $N$ spins each of magnitude $S$ in a 1D periodic cell.
By adopting the 1st NN interactions, the Heisenberg Hamiltonian is written as
\begin{equation}
\mathcal{H} = -2J \sum_{p=1}^N \mathbf{S}_p \cdot \mathbf{S}_{p+1}.
\end{equation}
Here $\hbar \mathbf{S}_p$ is the spin angular momentum at site $p$.
The interactions on the $p$th spin are
\begin{equation}
\mathcal{H}_p = -2J \mathbf{S}_p \cdot (\mathbf{S}_{p-1} + \mathbf{S}_{p+1})
              \equiv -\boldsymbol{\mu}_p \cdot \mathbf{B}_p,
\end{equation}
where $\boldsymbol{\mu}_p = -g \mu_\text{B} \mathbf{S}_p$ and $\mathbf{B}_p = -(2J/g\mu_\text{B}) (\mathbf{S}_{p-1} + \mathbf{S}_{p+1})$ are the magnetic moment and effective magnetic field at site $p$, respectively.
The time-dependent dynamics of $\hbar \mathbf{S}_p$ is defined by the torque $\boldsymbol{\mu}_p \times \mathbf{B}_p$ as
\begin{equation}
d\mathbf{S}_p/dt = (-g\mu_\text{B}/\hbar) \mathbf{S}_p \times \mathbf{B}_p.
\label{eq:torque}
\end{equation}

At the final step, the \textit{ab initio} calculations have been performed to realize our model analysis in AFM CrN.
The LDA \cite{Ceperley1980} plus $U$ approach \cite{Dudarev1998} is used with the projector augmented wave (PAW) potentials \cite{Kresse1999}, implemented in the \textsc{vasp} package \cite{Kresse1996a,*Kresse1996b}.
The crystal structure of AFM CrN (space group $Pnma$, No.\ 62) is fully optimized using $U = 4.1$~eV for the $3d$ electrons at Cr sites.
To explore the magnonic properties, a tight-binding model is constructed from the electronic structure using the \textsc{wannier90} library \cite{Mostofi2014}, including the Cr $3d$ and N $2p$ characters as a basis set.
The magnetic exchange parameters $J_{ij}$ are extracted for neighboring Cr pairs with indices $i$ and $j$ given in the Hamiltonian
\begin{equation}
\mathcal{H} = -2\sum_{i>j}J_{ij}\mathbf{S}_i\cdot\mathbf{S}_j,
\end{equation}
using the \textsc{tb2j} code \cite{He2021} within the local force theorem \cite{Liechtenstein1984}.
The magnon spectra are computed by applying linear spin wave theory that is a reliable approximation for systems with relatively large spin, including the present case with $S = 3/2$.
The algorithm \textsc{spinw} \cite{Toth2015} based on the Holstein-Primakoff approximation of spin operators is employed.
The classical Monte-Carlo simulations are carried out with the \textsc{vampire} software \cite{Evans2014} to obtain the temperature dependence of the AFM-ordered magnetic moments and derive the theoretical N\'{e}el temperature.

\section{Results and discussion}
\subsection{Model 1: Linear chain}
We start with the simple 1D model of double-layered antiferromagnets as illustrated in Fig.~\ref{fig:struct}(a).
The spins of $S^z = +S$ and $-S$ are indicated by the filled ($\bullet$) and open ($\circ$) circles, respectively.
The intra- and inter-sublattice interactions of $J_1$ and $J_2$ are implemented in Eq.~\eqref{eq:torque} according to the geometry of Fig.~\ref{fig:struct}(a).
The spin indices are $4p + n$, where $n = 0$, 1, 2, and 3 for the four magnetic sites in the $p$th unit cell.
The lattice parameter is $a = 4d$ with the same spacing $d$ for all sites.
Let us focus on the case of $n = 0$.
By assuming $S^{x,y} \ll S$, we get the linearized equations in Cartesian components
\begin{equation}
\begin{split}
dS_{4p}^x/dt &=  (2S/\hbar)(\Delta JS_{4p}^y - J_2S_{4p-1}^y - J_1S_{4p+1}^y), \\
dS_{4p}^y/dt &= -(2S/\hbar)(\Delta JS_{4p}^x - J_2S_{4p-1}^x - J_1S_{4p+1}^x),
\end{split}
\label{eq:torque_xy}
\end{equation}
with $\Delta J = J_1 - J_2$.
After repeating Eq.~\eqref{eq:torque_xy} for other three sites, the combined expressions using the ladder operator $S^+ = S^x + iS^y$ are
\begin{equation}
\begin{split}
dS_{4p  }^+/dt &=-i(2S/\hbar)(\Delta JS_{4p  }^+ - J_2S_{4p-1}^+ - J_1S_{4p+1}^+), \\
dS_{4p+1}^+/dt &=-i(2S/\hbar)(\Delta JS_{4p+1}^+ - J_1S_{4p  }^+ - J_2S_{4p+2}^+), \\
dS_{4p+2}^+/dt &= i(2S/\hbar)(\Delta JS_{4p+2}^+ - J_2S_{4p+1}^+ - J_1S_{4p+3}^+), \\
dS_{4p+3}^+/dt &= i(2S/\hbar)(\Delta JS_{4p+3}^+ - J_1S_{4p+2}^+ - J_2S_{4p+4}^+),
\end{split}
\label{eq:torque_ladder}
\end{equation}
where the solutions of $S^+$ are traveling spin waves
\begin{equation}
S_{4p+n}^+ = u_n \exp[i(4p+n)kd - i\omega t].
\label{eq:travel_wave}
\end{equation}
By substituting this into Eq.~\eqref{eq:torque_ladder}, four linear equations for the wavefunction coefficients $u_n$ are obtained and become solvable for a specific dispersion $\omega(k)$.
The magnon dispersion relation is acquired by solving the following determinant to be zero
\begin{align}
\begin{array}{|cccc|}
 \Delta \omega-\omega &-\omega_1 e^{ikd}     & 0                    &-\omega_2 e^{-ikd} \\
-\omega_1 e^{-ikd}    & \Delta \omega-\omega &-\omega_2 e^{ikd}     & 0                 \\
 0                    &-\omega_2 e^{-ikd}    & \Delta \omega+\omega &-\omega_1 e^{ikd}  \\
-\omega_2 e^{ikd}     & 0                    &-\omega_1 e^{-ikd}    & \Delta \omega+\omega
\end{array} = 0,
\end{align}
which leads to the secular equation
\begin{equation}
\omega^4 - 4(\omega_1^2 - \omega_1\omega_2)\omega^2 + 4\omega_1^2\omega_2^2 \sin^2(2kd) = 0,
\end{equation}
where $\omega_{1,2} = (2S/\hbar)J_{1,2}$ and $\Delta \omega = \omega_1 - \omega_2$.
We get the spectrum of acoustic ($-$) and optical ($+$) modes
\begin{equation}
\omega(k) = \omega_0\sqrt{A \pm B(k)},
\label{eq:omega}
\end{equation}
where $\omega_0 = \sqrt{2|\omega_1\omega_2|}$ is the characteristic frequency and
\begin{equation}
\begin{split}
A    ={}& (\sgn{\gamma})(\gamma - 1), \\
B(k) ={}& \sqrt{(\gamma - 1)^2 - \ssin^2(2kd)},
\end{split}
\end{equation}
with $\gamma = J_1/J_2$.
To ensure the stability of the $|{+}{+}{-}{-}\rangle$ ground state, any imaginary solutions should be avoided.
This gives the condition $\gamma < 0$, which means that alternation of positive $J_1$ and negative $J_2$ is required.

\begin{figure}[t]
\includegraphics[scale=0.34]{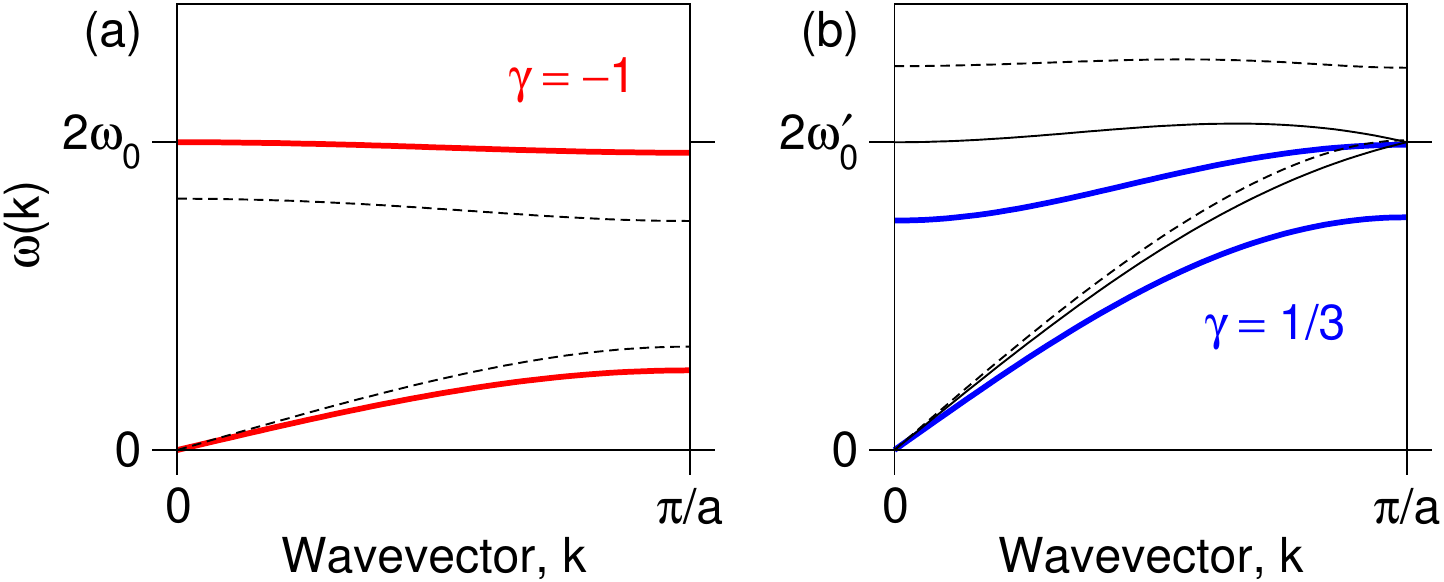}
\caption{
Selected magnon dispersion relations of double-layered antiferromagnets in the models of (a) linear chain and (b) railroad trestle (RT).
Four types of lines indicate $\gamma = J_1/J_2 = -1$ (red), $-1/3$ (black dashed), 0 (black solid), and $1/3$ (blue).
The characteristic frequencies are $\omega_0 = \sqrt{2|\omega_1\omega_2|}$ and $\omega_0' = \sqrt{2}|\omega_2|$.
Here positive $\gamma$ of $1/3$ means all magnetic interactions in RT being antiferromagnetic.
}
\label{fig:model}
\end{figure}

The plot of $\omega(k)$ in Eq.~\eqref{eq:omega} is shown in Fig.~\ref{fig:model}(a) for $\gamma = -1$ and $-1/3$.
Each of the acoustic and optical branches is 2-fold degenerate (see Appendix A for the illustration of spin wave modes and double degeneracy associated with chirality), hence there are in total four magnon bands.
A characteristic feature is the magnon gap at the Brillouin zone (BZ) boundary $k = \pi/a$.
The gap is calculated to $\sqrt{2}\omega_0$ for $\gamma = -1$, while it vanishes for $\gamma \rightarrow 0^-$.

\subsection{Model 2: Railroad trestle}
In solving the model with intra-sublattice interaction of $J_1$ and inter-sublattice interactions of $J_2$ and $J_3$ as illustrated in Fig.~\ref{fig:struct}(b), we now consider both the FM $J_1$ and the less obvious case of AFM $J_1$.
For simplicity, we take the other AFM parameters equal ($J_2 = J_3 < 0$).
After going through a more complex process than for the LC above, we get the spectrum (see Appendix B for the derivation of the corresponding secular equation)
\begin{equation}
\omega(k) = \omega_0' \sqrt{C(k) \pm D(k)},
\label{eq:omega_rt}
\end{equation}
where $\omega_0' = \sqrt{2}|\omega_2|$ is the characteristic frequency and 
\begin{equation}
\begin{split}
C(k) ={}& (\gamma -2)(\gamma - 1) + 2\sin^2(2kd), \\
D(k) ={}& \sqrt{(\gamma - 2)[\gamma^2(\gamma - 4) + (5\gamma - 2) \cos^2(2kd)]}.
\end{split}
\label{eq:cd}
\end{equation}
The spin-paired arrangement in Fig.~\ref{fig:struct}(b) is stable when $\omega(k)$ is real, which corresponds to $C(k) \pm D(k) \geq 0$ and real $D(k)$.
This is fulfilled for the range of $\gamma$ from any negative value (FM $J_1$) to a positive limit of $\gamma = 2/3$ (AFM $J_1$), i.e., in the interval $\gamma \leq 2/3$ ($J_1 \geq -2|J_2|/3$).
There is a finite magnon gap at the BZ boundary, except for $\gamma = 0$.

\begin{figure}[t]
\includegraphics[scale=0.34]{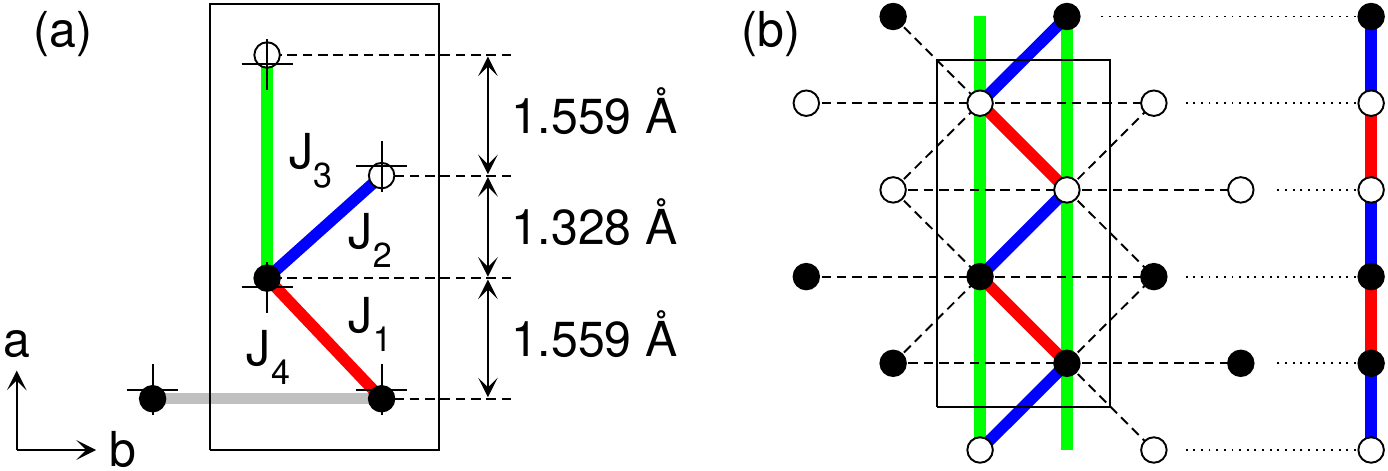}
\caption{
(a) The crystal structure and magnetic ordering of CrN projected on the orthorhombic $ab$ plane.
The symbols ``$\bullet$ ($\circ$)'' and ``$+$'' indicate the Cr sites of $S^z = +S$ ($-S$) and N sites, respectively.
Four kinds of magnetic exchange parameters $J_1$ (red), $J_2$ (blue), $J_3$ (green), and $J_4$ (grey) are marked.
For more detailed structural description, see Ref.~\cite{Ahn2024}.
(b) Schematic two-dimensional picture of exchange links.
The full lines denote the motif running along the $a$ axis that closely reminds the railroad trestle in Fig.~\ref{fig:struct}(b) with magnetic exchange parameters $J_1$, $J_2$, and $J_3$.
The dashed lines show links toward Cr atoms in neighboring units that include in addition to $J_1$ and $J_2$, also a new parameter $J_4$.
The projection to the $a$ axis, seen on the right side, refers to the spin-paired linear chain of Fig.~\ref{fig:struct}(a).
}
\label{fig:crn}
\end{figure}

To illustrate the magnon spectra, we used the special case of $\gamma = 0$, for which Eq.~\eqref{eq:omega_rt} can be simplified.
The solution reduces to
\begin{equation}
\omega(k) =
\begin{cases}
2\omega_0' |\sin kd| \sqrt{3 - 2\ssin^2kd} & (\text{Acoustic}), \\
2\omega_0' |\cos kd| \sqrt{1 + 2\ssin^2kd} & (\text{Optical}).
\end{cases}
\label{eq:omega_g0}
\end{equation}
The dispersion relations of Eqs.~\eqref{eq:omega_rt} and \eqref{eq:omega_g0} are presented in Fig.~\ref{fig:model}(b) for $\gamma = -1/3$, 0, and $1/3$.

\begin{table*}[t]
\caption{
Summary of the distances $d_{ij}$ and magnetic exchange parameters $J_{ij}$ among the four kinds of nearest-neighbor (NN) $i$--$j$ pairs, from the \textit{ab initio} calculations with $U = 4.1$~eV.
We consider three variants of the CrN structure by keeping the same volume of the magnetic unit cell.
(i) The hypothetical cubic phase of rock-salt (RS) type.
(ii) The intermediate structure upon orthorhombic distortion but without (w/o) local atomic shifts (AS).
(iii) The real fully optimized orthorhombic phase with (w/) AS.
The lattice parameters are $(a, b, c) = (\sqrt{2}c, c/\sqrt{2}, c) = (5.8536, 2.9268, 4.1391)$~\AA{} for (i) and $(a, b, c) = (5.7743, 2.9647, 4.1421)$~\AA{} for (ii,~iii).
The atomic positions are specified by $(x_\text{Cr}, 1/4, 1/4)$ for the Cr and $(x_\text{N}, 1/4, 3/4)$ for the N sites, where $(x_\text{Cr}, x_\text{N}) = (1/8, 1/8)$ for (i,~ii) and $(x_\text{Cr}, x_\text{N}) = (0.1350, 0.1146)$ for (iii).
The intra-sublattice (SL) interaction of index $J_4$ arises in the three-dimension, connecting the neighboring unit cells in the $b$ direction.
The full list of the distance vectors is given by $\mathbf{d}_{ij} = \pm d_{ij}^x\hat{x}\ {\pm(\pm)(\mp)(\mp)}\ d_{ij}^y\hat{y}\ {\pm(\mp)(\pm)(\mp)}\ d_{ij}^z\hat{z}$.
}
\label{table:param}
\begin{ruledtabular}
\begin{tabular}{ccccccccccc}
Type & Index & \multicolumn{3}{c}{Distance vector $\mathbf{d}_{ij}$} & \multicolumn{2}{c}{(i) Cubic RS} & \multicolumn{2}{c}{(ii) Ortho w/o AS} & \multicolumn{2}{c}{(iii) Ortho w/ AS} \\
\cline{3-5} \cline{6-7} \cline{8-9} \cline{10-11}
& & ${d}_{ij}^x$ & ${d}_{ij}^y$ & ${d}_{ij}^z$ & $|d_{ij}|$ (\AA) & $J_{ij}$ (meV) & $|d_{ij}|$ (\AA) & $J_{ij}$ (meV) & $|d_{ij}|$ (\AA) & $J_{ij}$ (meV) \\
\hline
Intra-SL & $J_1$ &         $2x_{\rm Cr}a$ & $b/2$ & $c/2$ & 2.9268 & $-$2.6361 & 2.9275 & $-$2.5951 & 2.9862 & $-$0.2553 \\
         & $J_4$ &                      0 &   $b$ &     0 & 2.9268 & $-$2.4309 & 2.9647 & $-$1.1701 & 2.9647 & $-$1.0128 \\
\hline
Inter-SL & $J_2$ & $(1/2 - 2x_{\rm Cr})a$ & $b/2$ & $c/2$ & 2.9268 & $-$2.9943 & 2.9275 & $-$2.9516 & 2.8724 & $-$4.9555 \\
         & $J_3$ &                  $a/2$ &     0 &     0 & 2.9268 & $-$2.6247 & 2.8872 & $-$3.7028 & 2.8872 & $-$3.5406
\end{tabular}
\end{ruledtabular}
\end{table*}

Here we summarize our model analyses.
The double-layered antiferromagnetism in LC is stabilized by alternating FM and AFM exchange parameters $(J_1, J_2) = ({+|J_1|}, {-|J_2|})$ with no restriction on the magnitudes of $J_1$ and $J_2$.
An analogous magnetic arrangement in the RT geometry involves three exchange parameters of an obvious combination $(J_1, J_2, J_3) = ({+|J_1|}, {-|J_2|}, {-|J_3|})$, but a negative sign of $J_1$ is also permitted.
In particular for $J_2 = J_3$, the stability of the double-layered antiferromagnetism with a $|{+}{+}{-}{-}\rangle$ unit cell, seen in Fig.~\ref{fig:struct}(b), is ensured down to $J_1 = -2|J_2|/3$.
As $J_1$ becomes more negative, a region of canted spin arrangements emerges and finally for $J_1 \leq -2|J_2|$, a distinct double-layered AFM phase with a $|{+}{-}{-}{+}\rangle$ unit cell becomes stable (see Appendix C).
Regarding magnetic excitations, a fundamental property of 1D double-layered antiferromagnets is the presence of both acoustic and optical branches, exhibiting a magnon gap at the BZ boundary.

\subsection{Antiferromagnetic CrN}
To investigate the stability of the observed double-layered AFM arrangement in CrN \cite{Corliss1960,Gui2022}, we performed \textit{ab initio} calculations of exchange interactions, incorporating both the macroscopic lattice distortion and local atomic displacements.
As depicted in Fig.~\ref{fig:crn}(a), the projection of AFM CrN on the orthorhombic $ab$ plane demonstrates two oppositely oriented magnetic sublattices, each formed by double $bc$ layers of Cr sites with a spacing of 1.559~\AA, while the inter-sublattice spacing is slightly smaller at 1.328~\AA{} \cite{Ahn2024}.
Fig.~\ref{fig:crn}(b) reveals the exchange links in AFM CrN within a two-dimensional (2D) scheme, highlighting its similarity to the 1D model of RT treated above.
In this orthorhombically distorted rock-salt structure, Cr atoms form a lattice close to a face-centered cubic arrangement, resulting in twelve NNs: four in a neighboring layer of same spin ($J_1$), four in the opposite neighboring layer of reversed spin ($J_2$), two in the next-nearest layers ($J_3$), and two within the same layer ($J_4$).
Important values of exchange parameters are found for these NN Cr pairs ($d \approx 2.9$~\AA), whereas interactions between more distant Cr pairs including the 180\textdegree{} Cr--N--Cr superexchange ($d \approx 4.1$~\AA) are negligible.

\begin{figure}[t]
\includegraphics[scale=0.34]{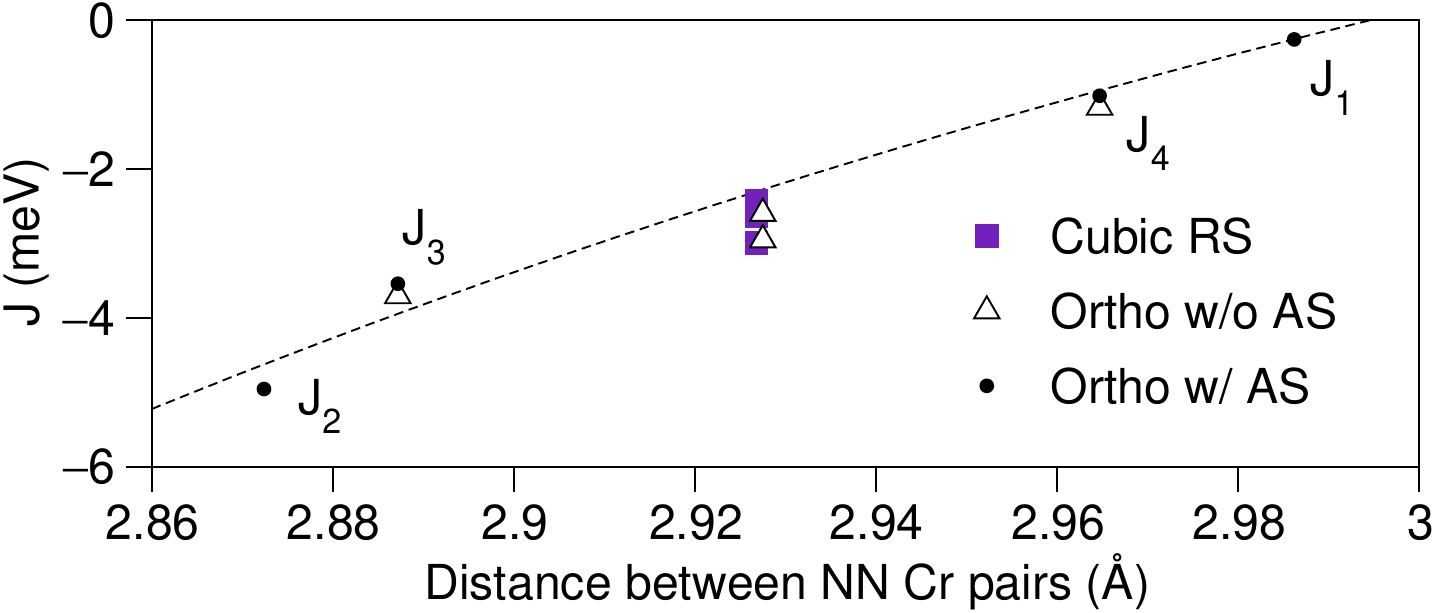}
\caption{
Calculated magnetic exchange parameters for three variants of the CrN structure: the hypothetical cubic phase of rock-salt (RS) type (purple squares), the intermediate structure upon orthorhombic distortion but without (w/o) local atomic shifts (AS) (empty triangles), and the real fully optimized orthorhombic phase with (w/) AS (black dots with $J_1$, \dots, $J_4$ denoted).
The dependence of the last data on the distance between nearest-neighbor (NN) Cr pairs of $S = 3/2$ is fitted as an interplay of Cr--Cr direct exchange and 90\textdegree{} Cr--N--Cr superexchange (dashed line, see the text).
}
\label{fig:dist}
\end{figure}

We considered three structural variants: (i) the hypothetical cubic rock-salt structure, (ii) the macroscopically distorted one, and (iii) the real CrN structure including local atomic shifts.
Table~\ref{table:param} summarizes the results, and Fig.~\ref{fig:dist} illustrates the correlation between exchange interactions and Cr--Cr distances.
Concerning variant (i), it is worth mentioning that all twelve NN distances are identical due to the strict cubic symmetry, but the exchange parameters are influenced by the double-layered AFM arrangement accordingly.
While $J_2$ and $J_3$ couple to antiparallel spin pairs, $J_1$ and $J_4$ refer to parallel spin pairs despite their AFM character, as seen in Fig.~\ref{fig:crn}(a).
The uniform size of $-$3~meV for $J_1$, \dots, $J_4$ may signify that the spin exchange can be effectively described by a Heisenberg-type Hamiltonian and would lead to highly frustrated magnetic interactions.

In contrast, variants (ii) and (iii) exhibit a gradual diversification of NN Cr--Cr distances, leading to a significant difference between the inter-sublattice exchanges $(J_2, J_3) = (-4.96, -3.54)~\text{meV} \ll 0$ and intra-sublattice ones $(J_1, J_4) = (-0.26, -1.01)~\text{meV} \lesssim 0$, ensuring that all remain negative.
Notably, the shortening of NN distances by $\sim$2\% enhances the AFM strength of $J_2$ from $-$3~meV to $-$5~meV, while a similar elongation induces some FM tendency and substantially reduces the AFM strength of $J_1$ from $-3$~meV to $-0.26$~meV.
We attribute this behavior to an interplay of two competing NN interactions that can be expressed by means of hopping integrals $t$.
These are the Cr--Cr direct exchange via $t_{2g}$--$t_{2g}$ bonding of $3d$ orbitals ($J_\text{AFM} \propto -t_{dd}^2$) and 90\textdegree{} Cr--N--Cr superexchange ($J_\text{FM} \propto t_{pd\sigma}^2t_{pd\pi}^2$), the latter arising from virtual hopping from the occupied $t_{2g}$ shell of one Cr to the empty $e_g$ shell of its NN Cr site, mediated by N $2p$ orbitals (see Appendix D for a detailed description of chemical bonds).
Regarding variant (iii), we modeled their combined effect with the following expression:
\begin{equation}
J = -\kappa d^{-10} + \lambda s^{-14},
\label{eq:fit}
\end{equation}
where $\kappa$ and $\lambda$ are the fitting parameters, and $s_{(1,2,3,4)}$ represents the Cr--N distances averaged over Cr--N--Cr paths that contribute to the given $J_{(1,2,3,4)}$ links.
In distinction to the $\sim$4\% difference in $d$ values, $s$ manifests only negligible variance ($\sim$0.15\%), allowing it to be approximated as a constant, $s = 2.07$~\AA, in Eq.~\eqref{eq:fit}.
The power-law dependence follows from known interatomic overlaps of two wavefunctions: $t_{dd} \propto d^{-5}$, $t_{pp} \propto d^{-2}$, and typically $t_{pd\sigma} \simeq \sqrt{2}t_{pd\pi} \propto d^{-3.5}$ \cite{Harrison1989}.
We fitted the data using $\kappa = 5.194 \times 10^5$~meV\,\AA$^{10}$ and $\lambda = 2.376 \times 10^5$~meV\,\AA$^{14}$, as indicated by the dashed line in Fig.~\ref{fig:dist}.
The calculated components read $J_\text{AFM} = -13.58$~meV for $J_2$ ($-9.21$~meV for $J_1$) and $J_\text{FM} = 8.96$~meV.

Note that a similar behavior, but with opposite tendencies (namely, the strengthening of FM interactions upon the shortening of NN distances at high pressures) has been found in mixed-valent perovskite manganites, where the interplay between Mn$^{3+}$--O--Mn$^{4+}$ double exchange (FM, $d^{-3.5}$) and 180\textdegree{} Mn$^{3+}$--O--Mn$^{3+}$ superexchange (AFM, $d^{-14}$) governs the exchange mechanism \cite{Pai2001,Kozlenko2004}.
An even more complex situation is encountered in elemental $3d$ metals like Fe, Co, and Ni, where despite their simple crystal structure there is an intricate interplay of different exchange interactions ($t_{2g}$--$t_{2g}$, $e_g$--$e_g$, and $t_{2g}$--$e_g$), some of them leading to non-Heisenberg terms, see Ref.~\cite{Kvashnin2016,Szilva2023}.

\begin{figure}[t]
\includegraphics[scale=0.34]{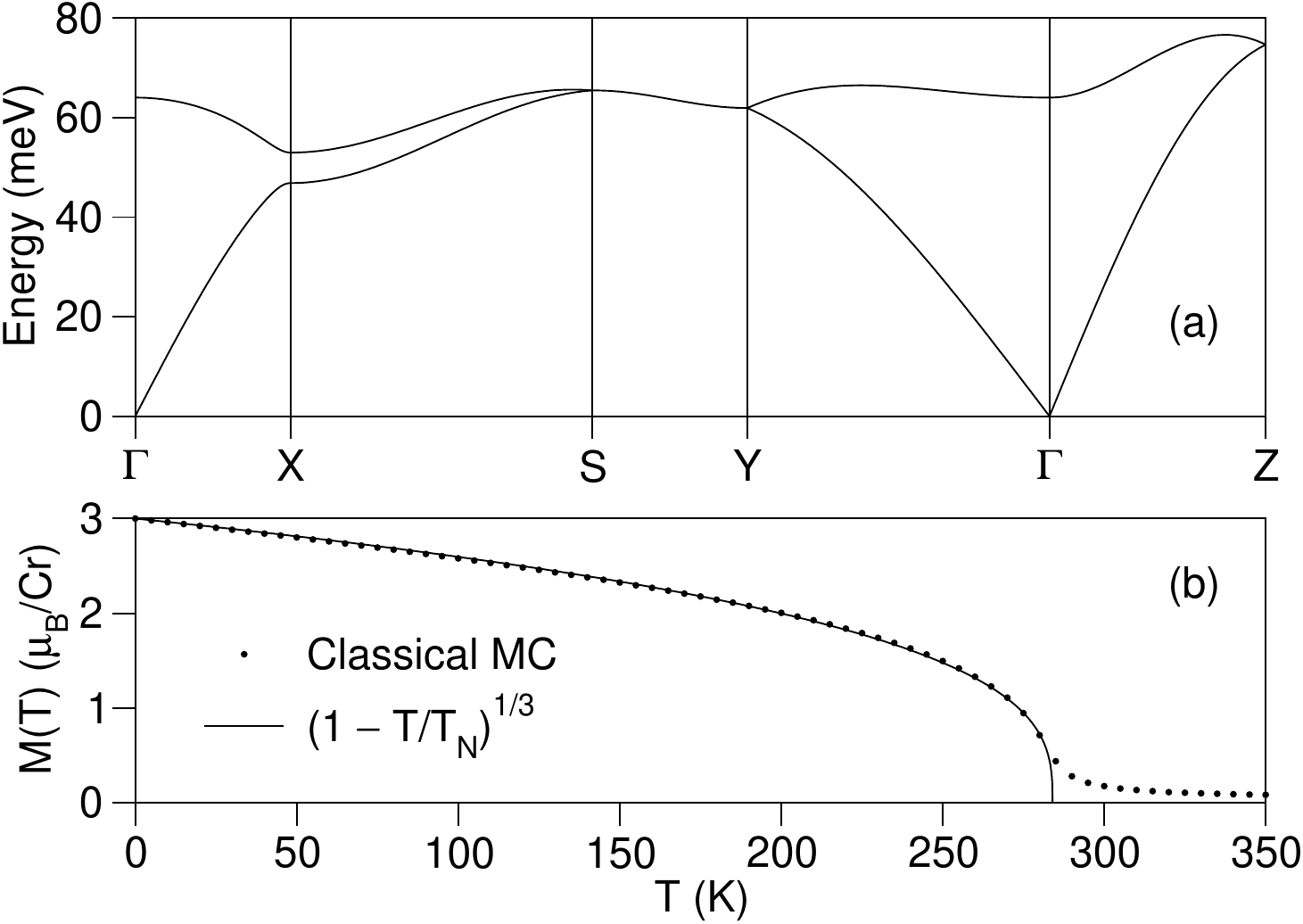}
\caption{
(a) Magnon dispersion relation of antiferromagnetic CrN.
The presence of a magnon gap at the X point shows good agreement with the model analysis in Fig.~\ref{fig:model}.
(b) Calculated temperature evolution of local magnetic moments from the classical Monte-Carlo simulations.
The ordering temperature of $T_\text{N} = 284$~K was obtained by fitting with $(1 - T/T_\text{N})^{1/3}$.
}
\label{fig:band}
\end{figure}

To calculate the magnon spectrum of AFM CrN, we used the \textit{ab initio} determined $J_1$, \dots, $J_4$ for constructing the magnetic Hamiltonian and subsequent transformation of spin excitations into bosonic operators.
The resulting dispersion relations for variant (iii) are presented in Fig.~\ref{fig:band}(a).
Since double-layered AFM arrangement holds along the crystallographic $a$ direction while the FM order is maintained along the $b$ and $c$ directions, the magnon gap appears only at the X point but not at the Y or Z points.
For the acoustic magnon branch, the linear (AFM-like) character of the dispersion is obvious.
Although it bears certain similarity to the acoustic phonons \cite{Ahn2024}, the slope of the dispersion relation is different, implying the absence of magnon-phonon coupling at low temperatures.
Fig.~\ref{fig:band}(b) shows the temperature dependence of local magnetic moments, obtained for the fixed orthorhombic structure of CrN.
The N\'{e}el temperature is determined to be $T_\text{N} = 284$~K, which is consistent with the reported experimental value of 287~K \cite{Wang2023}, confirming the validity of ours $J$.
Also, the largest value found for $J_2$ is in good agreement with Ref.~\cite{Biswas2023} ($J_2S^2 = -11.15$~meV).

A final comment should be given to the fact that the AFM transition has been realistically calculated without considering the eventual role of concomitant transition to the cubic structure.
It should be noted that two CrN phases, the AFM orthorhombic and paramagnetic cubic ones, are found dynamically stable by using both the density functional theory based approximation (see e.g., the phonon spectra in Fig.~2 of Ref.~\cite{Ahn2024}) and the more advanced approach that takes into account dynamic coupling of vibrational states to disordered spins and is valid for finite temperatures (see Figs.~1--3 of Ref.~\cite{Shulumba2014}; a more detailed information on the disordered moment method can be found in Refs.~\cite{Lindmaa2013,Stockem2018}).
The magnetostructural transition in CrN is thus of first order---there is no gradual evolution toward the cubic structure in the low-temperature region; instead, once the AFM order sharply collapses at $T_\text{N}$ and the gain in exchange energy vanishes, the cubic phase, which possesses much higher vibrational and magnetic entropy, becomes thermodynamically favorable.

\section{Conclusions}
We studied the stability and magnonic properties of double-layered AFM order using 1D models and validated our findings with \textit{ab initio} calculations for CrN.
We emphasize that the LC and RT geometries in Fig.~\ref{fig:struct} serve as minimal yet ideal models for exploring the interplay between $J_1$ and $J_2$, as both the intra- ($\multimapdotboth$,~$\multimapboth$) and inter-sublattice ($\multimapdotbothB$) exchanges are involved at every site.
A key distinction between the LC and RT models lies in the ratio of parallel to antiparallel spin pairs, $1:1$ (LC) and $1:3$ (RT), which modifies the allowed range of the intra-sublattice interaction from $J_1 > 0$ (LC) to $J_1 > -2|J_2|/3$ (RT).
Furthermore, we identify a unique case in which a fully 3D system (CrN) can be effectively captured by a 1D model (RT), evidenced by all $J_1$, \dots, $J_4$ being negative and by the presence of a magnon gap at the BZ boundary.
This applicability arises from the dominant target $|{+}{+}{-}{-}\rangle$ order, which is confined to a single crystallographic direction in double-layered antiferromagnets.

Concerning specific features of CrN, we note that the stability of its double-layered AFM order is intimately associated with the orthorhombic structural distortion, which leads to a noticeable diversification of exchange interactions between NN Cr sites.
As seen in Fig.~\ref{fig:crn}(a), there are two types of exchange parameters for intra-sublattice pairs and two types for inter-sublattice ones, and the ratio of parallel to antiparallel spin pairs is $6:6$ for every Cr site.
In the \textit{ab initio} calculations, we find strong AFM inter-sublattice parameters $J_{2,3} \ll 0$ and weaker but still of AFM character the intra-sublattice parameters $J_{1,4} \lesssim 0$.
Importantly, the $J_1$, \dots, $J_4$ values show a clear correlation with the corresponding Cr--Cr distances, which we attribute to the competition between Cr--Cr direct exchange (AFM) and 90\textdegree{} Cr--N--Cr superexchange (FM) at each NN link.
The exchange parameters obtained have been further used for calculation of the temperature dependence of magnetic ordering.
Considering the present results and some supporting data published elsewhere, we argue for the transition from the AFM orthorhombic to paramagnetic cubic CrN phase, occurring at 287 K, to be of the first-order character.

\begin{acknowledgments}
S.-J.\ K.\ gratefully acknowledges studentship support from the International Max Planck Research School for Chemistry and Physics of Quantum Materials.
This work was supported by Grants No.\ 22-10035K, No.\ 23-04746S, and No.\ 25-17490S of the Czech Science Foundation and Grant No.\ 471878653 of the German Research Foundation.
Computational resources were provided by e-INFRA CZ Grant No.\ 90254.
We further acknowledge the Operational Program Research, Development and Education financed by the European Structural and Investment Funds and by the Ministry of Education, Youth and Sports of the Czech Republic, Grant No.\ CZ.02.01.01/00/22\_008/0004594 (TERAFIT).
\end{acknowledgments}

\section*{Data availability}
Data associated with this study are available on the Zenodo repository \cite{Kim2024}.

\appendix

\section{Spin waves in the linear chain model}
\begin{figure}[t]
\includegraphics[scale=0.34]{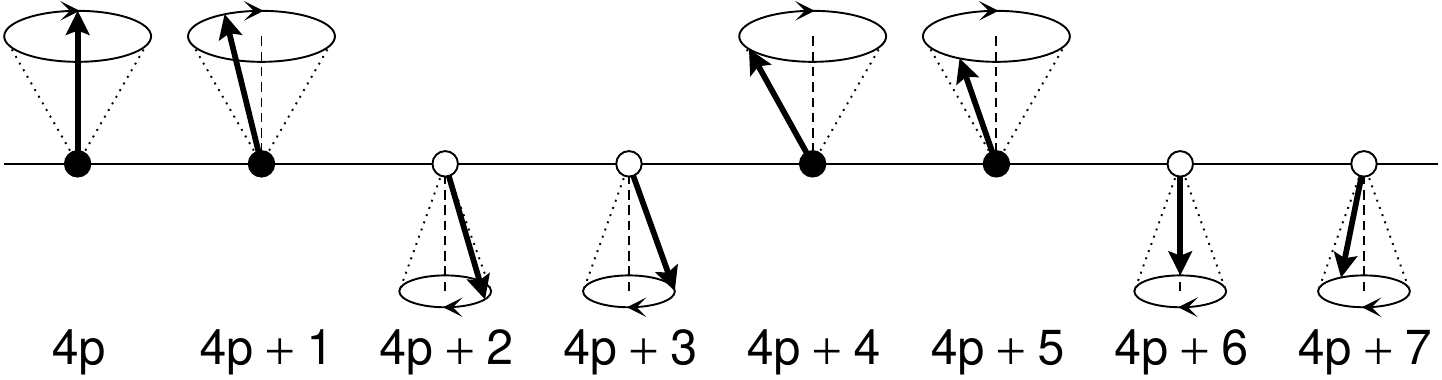}
\caption{
Semiclassical picture of an acoustic magnon in the linear chain of $|{+}{+}{-}{-}\rangle$ order.
Here the phase shifts between the nearest-neighbor sites are set to $kd = 30$\textdegree{} with $\alpha = \beta = 0$ for better visibility.
The bold vectors show an instantaneous orientation of atomic spins for the magnon propagating along the horizontal right direction with positive $k$.
}
\label{fig:wave}
\end{figure}

Within the present approximation valid for $S^{x,y} \ll S$, the spin dynamics of the LC with collinear $|{+}{+}{-}{-}\rangle$ order seen in Fig.~\ref{fig:struct}(a) is described by Eqs.~\eqref{eq:torque_ladder} and \eqref{eq:travel_wave}.
The solution for a given reciprocal vector $k$ gives four traveling spin waves of the conical spiral form: two degenerate acoustic modes and two degenerate optical ones.
An example of the acoustic mode is schematically shown in Fig.~\ref{fig:wave}.
This magnon is characterized by larger cone angles for spin-up sites and clockwise rotations of local spins (positive $\omega$), while its degenerate counterpart possessing opposite chirality would display exactly reversed cone angles and counter-clockwise rotation of atomic spins (negative $\omega$).
Such a situation is analogous to that discussed in detail for the simple $|{+}{-}{+}{-}\rangle$ chain by Keffer \textit{et al}.\ \cite{Keffer1953}.
We underline that, in the present $|{+}{+}{-}{-}\rangle$ chain with four magnetic sites $n = 0$, 1, 2, and 3, the calculated spin wave parameters $u_0$, $u_1$, $u_2$, and $u_3$ are generally of complex values.
This implies that phase shifts $\varphi$ between neighboring sites are not regular ($\varphi = kd$) as Eq.~\eqref{eq:travel_wave} seems to suggest, but they are diversified for the spin-up pairs, spin-down pairs, and spin-antiparallel pairs.

\begin{figure}[t]
\includegraphics[scale=0.34]{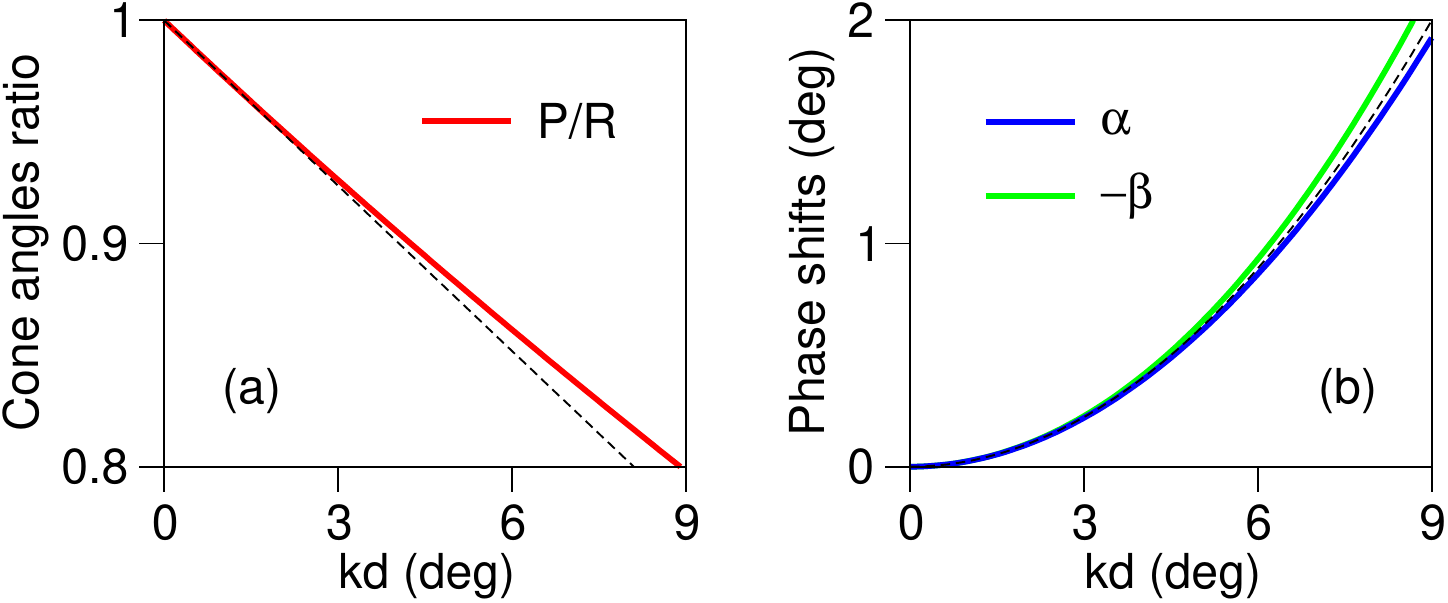}
\caption{
Spin wave parameters: $P/R$ (red), $\alpha$ (blue), and $-\beta$ (green) in the small $k$ regime up to $kd = 9$\textdegree, corresponding to $k = 0.2\pi/a$.
The dashed lines in (a) and (b) represent the leading terms, $1 - \sqrt{2}kd$ and $\sqrt{2}(kd)^2$, respectively.
}
\label{fig:phase}
\end{figure}

Fig.~\ref{fig:phase} summarizes the spin wave parameters calculated for the selected case of $\gamma = -1$, expressed suitably in a polar form $u_{0(1)} = R\exp[+(-)i\alpha/2]$ and $u_{2(3)} = -P\exp[+(-)i\beta/2]$.
It appears that the ratio of the spin-down and spin-up cone angles decreases with increasing $k$ as $P/R \simeq 1 - \sqrt{2}kd + (kd)^2$.
The additional phase shifts $\alpha$ ($\beta$) between spin-up (spin-down) pairs have opposite signs and show an essentially quadratic dependence on $k$ as $\alpha~(\beta) \simeq +(-)\sqrt{2}(kd)^2 - (kd)^3/2$.
The actual phase shift between spin-up pairs is thus decreased (to $\varphi = kd - \alpha$), while that between spin-down pairs is similarly increased.

\section{Equations of motion for the spin chain of railroad trestle type}
We consider the dynamics of the RT model in Fig.~\ref{fig:struct}(b).
In distinction to Eq.~\eqref{eq:torque_ladder} for the simple LC, the torque acting on spins in RT includes two more terms due to a larger number of NN spins.
There are again four equations of spin motion in local ladder operators
\begin{equation}
\begin{split}
dS_{4p}^+/dt   &= -i(2S/\hbar)[\Delta J S_{4p}^+   - J_2 S_{4p-1}^+ - J_1 S_{4p+1}^+ \\
      &\phantom{= -i(2S/\hbar)[} - J_3(S_{4p-2}^+ + S_{4p+2}^+)], \\
dS_{4p+1}^+/dt &= -i(2S/\hbar)[\Delta J S_{4p+1}^+ - J_1 S_{4p}^+   - J_2 S_{4p+2}^+ \\
      &\phantom{= -i(2S/\hbar)[} - J_3(S_{4p-1}^+ + S_{4p+3}^+)], \\
dS_{4p+2}^+/dt &=  i(2S/\hbar)[\Delta J S_{4p+2}^+ - J_2 S_{4p+1}^+ - J_1 S_{4p+3}^+ \\
      &\phantom{=  i(2S/\hbar)[} - J_3(S_{4p}^+   + S_{4p+4}^+)], \\
dS_{4p+3}^+/dt &=  i(2S/\hbar)[\Delta J S_{4p+3}^+ - J_1 S_{4p+2}^+ - J_2 S_{4p+4}^+ \\
      &\phantom{=  i(2S/\hbar)[} - J_3(S_{4p+1}^+ + S_{4p+5}^+)],
\end{split}
\label{eq:torque_ladder_rt}
\end{equation}
where $\Delta J = J_1 - J_2 - 2J_3$ and the same solutions for $u_n$ with Eq.~\eqref{eq:travel_wave}.
The magnon dispersion relation is obtained by solving
\begin{align}
\begin{array}{|cccc|}
  \Delta \omega-\omega & -\omega_1 e^{ikd}     &f_3(k)     & -\omega_2 e^{-ikd} \\
 -\omega_1 e^{-ikd}    &  \Delta \omega-\omega & -\omega_2 e^{ikd}      &f_3(k) \\
f_3(k)    & -\omega_2 e^{-ikd}    &   \Delta \omega+\omega & -\omega_1 e^{ikd} \\
 -\omega_2 e^{ikd}     &f_3(k)    & -\omega_1 e^{-ikd}     &  \Delta \omega+\omega
\end{array} = 0,
\label{eq:determinant_rt}
\end{align}
where $\omega_{1,2,3} = (2S/\hbar)J_{1,2,3}$, $\Delta \omega = \omega_1 - \omega_2 - 2\omega_3$, and $f_3(k) = -2\omega_3\cos(2kd)$.
An analytical solution is possible only for specific combinations of $J_1$, $J_2$, and $J_3$, as exemplified by Eqs.~\eqref{eq:omega_rt} and \eqref{eq:cd} for $J_2 = J_3$.

\section{Stability of collinear antiferromagnetic states in the railroad trestle model}
Taking into account the magnetic frustration arising when all exchange parameters $J_1$, $J_2$, and $J_3$ are negative in the RT model, we investigate the stability of three collinear spin arrangements: the discussed double-layered AFM2 with a four-atom $|{+}{+}{-}{-}\rangle$ unit cell seen in Fig.~\ref{fig:struct}(b), its spin-modified counterpart AFM2$'$ with a $|{+}{-}{-}{+}\rangle$ unit cell, and the single-layered AFM1 with a $|{+}{-}{+}{-}\rangle$ unit cell.
Their respective energies are given by $E_\text{AFM2} = 4(-J_1+J_2+2J_3)S^2$, $E_{\text{AFM2}'} = 4(J_1-J_2+2J_3)S^2$, and $E_\text{AFM1} = 4(J_1+J_2-2J_3)S^2$.
The conventional spin alternation of AFM1 is stabilized when the strengths of negative-valued $J_1$ and $J_2$ dominate over the $J_3$ one.
Our interest is, nonetheless, in the special case of $J_2 = J_3 < 0$, for which the RT model gives results similar to behavior of the real CrN.
It is obvious that for $J_1 \geq 0$ ($\gamma \leq 0$), there is no frustration, and the collinear $|{+}{+}{-}{-}\rangle$ arrangement is the real ground state; intuitively, one may expect it stable even for some negative and small $J_1$.
Clear frustration occurs at $J_1 = J_2$, where $E_\text{AFM2} = E_{\text{AFM2}'}$.
For more negative $J_1 < -|J_2|$ ($\gamma > 1$), the AFM2$'$ with the shifted $|{+}{-}{-}{+}\rangle$ order has the lowest energy among the three.

\begin{figure}[t]
\includegraphics[scale=0.34]{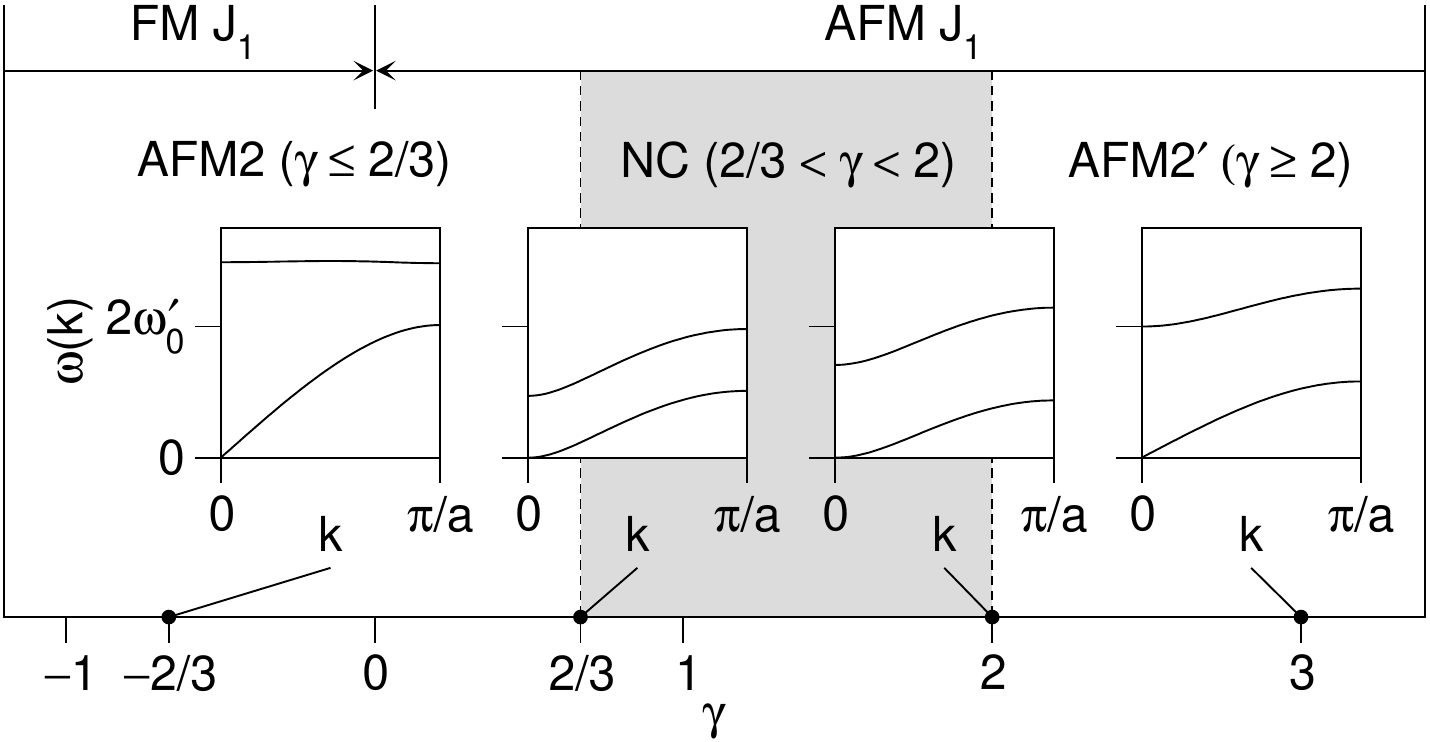}
\caption{
The stability of collinear spin arrangements of the railroad trestle.
Exchange parameters $J_2$ and $J_3$ are considered to be antiferromagnetic (AFM) and equal ($J_2 = J_3 < 0$), while $J_1$ varies from ferromagnetic (FM) to AFM.
Lower axis denotes the values of $\gamma = J_1/J_2$.
The region of AFM2 with a $|{+}{+}{-}{-}\rangle$ unit cell is characterized by magnon spectra for $\gamma = -2/3$ and $2/3$ (see Fig.~\ref{fig:model}(b) for $\gamma = -1/3$, 0, and $1/3$), and that of AFM2$'$ with a $|{+}{-}{-}{+}\rangle$ unit cell by spectra for $\gamma = 2$ and 3.
The characteristic frequency is $\omega_0' = \sqrt{2}|\omega_2|$.
In the middle part, the calculations of acoustic branches for AFM2/AFM2$'$ magnons give non-physical solutions, suggesting a region of non-collinear (NC) ground states.
}
\label{fig:range}
\end{figure}

Considering the possibility of non-collinear arrangements, we further examine the stability conditions of the AFM2 and AFM2$'$ states.
We then shift our focus from $E(J, S)$ to $\omega(k)$.
Fig.~\ref{fig:range} shows the calculated magnon spectra for both arrangements at $J_2 = J_3 < 0$.
In the range $2/3 < \gamma < 2$, the results reveal imaginary solutions in some part of the acoustic magnon branch, indicating the instability of both the $|{+}{+}{-}{-}\rangle$ and $|{+}{-}{-}{+}\rangle$ configurations and suggesting non-collinear ground states.
On the other hand, we confirm that the AFM2 arrangement of current interest represents a ground state for any positive $J_1$ ($\gamma < 0$) and also for small negative $J_1$ ($0 \leq \gamma \leq 2/3$).
The stability region for the alternative AFM2$'$ phase is found for negative $J_1$ dominating over $J_2$ and $J_3$ ($\gamma \geq 2$).

It should be noted that the present semiclassical approach to AFM ground states and the magnon dynamics is appropriate only for atomic moments with large spin values.
For small spins, in particular $S = 1/2$, a more complex quantum mechanical solution is needed.
Spin arrangements of the N\'{e}el type have thus been shown not to be true eigenstates of the isotropic Heisenberg Hamiltonian with an AFM exchange parameter.
The solutions for LCs with periodic boundary conditions can be found in the instructive paper by Bonner and Fisher \cite{Bonner1964}.
By taking different N\'{e}el configurations (alternations of atomic spins $S^z = \pm1/2$) as basis vectors, it is shown that AFM eigenstates are linear combinations of all basis vectors with zero total spin, irrespective of the unit-cell length.
For the four-atom unit cell, the ground state for simple linear AFM chain is formed by a linear combination of $|{+}{-}{+}{-}\rangle$, $|{-}{+}{-}{+}\rangle$, $|{+}{+}{-}{-}\rangle$, $|{-}{-}{+}{+}\rangle$, $|{+}{-}{-}{+}\rangle$, and $|{-}{+}{+}{-}\rangle$.
Findings for variable unit cells (up to $N = 12$) demonstrate the absence of long-range magnetic order in the AFM chain.
The extrapolated energy of the ground state corresponds to the exact value, $-2N|J|(\ln{2} - 1/4) = -0.88629N|J|$, derived in the famous paper by Bethe \cite{Bethe1931}.
Compared to the FM chain, which has a quantum mechanical energy of $-2NJS^2$ for any $S$ value, the energy of AFM chain for $S = 1/2$ can be formally written as $-2N|J|S(S + 0.38629)$.

\begin{figure}[t]
\includegraphics[scale=0.34]{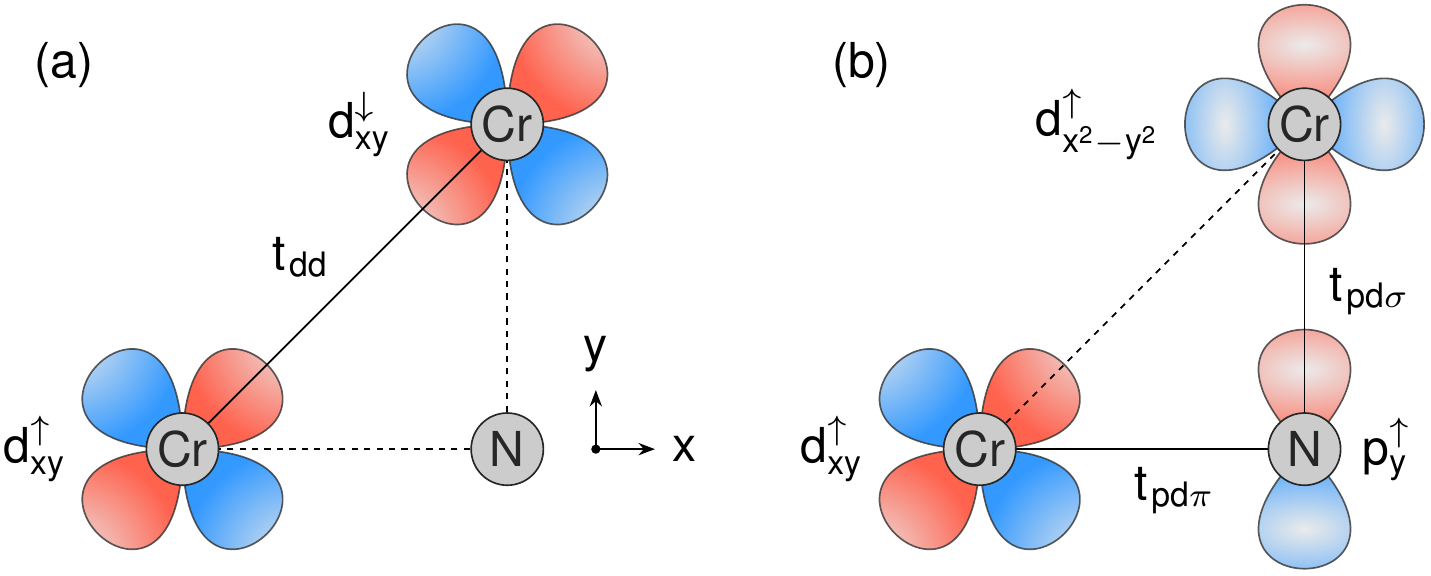}
\caption{
Orbital links decisive for the (a) antiferromagnetic Cr--Cr direct exchange and (b) ferromagnetic 90\textdegree{} Cr--N--Cr superexchange.
}
\label{fig:orbit}
\end{figure}

\section{Chemical bonds in CrN}
A strong correlation of the $J_1$, \dots, $J_4$ values with the corresponding Cr--Cr distances in AFM CrN is evident in Fig.~\ref{fig:dist}.
The origin of this behavior can be traced to the nature of Cr--Cr and Cr--N chemical bonds, in which the spin polarization of bonding electrons plays an essential role.
As illustrated in Fig.~\ref{fig:orbit} (see also the analogous orbital schemes in Fig.~5 of Ref.~\cite{Ushakov2013}), the first factor decisive for the exchange interaction is the overlap of an occupied Cr $t_{2g}^\uparrow$ orbital with an unoccupied $t_{2g}^\uparrow$ orbital at its NN, which leads to AFM Cr--Cr direct exchange.
On the other hand, the same Cr pair is also connected by two different 90\textdegree{} Cr--N--Cr links.
In the Cr--N direction, there is an overlap of a Cr $t_{2g}^\uparrow$ orbital with an N $p^\uparrow$ orbital ($\pi$ bonding), which results in strong hybridization and consequently, in a different spatial distribution for the $p^\uparrow$ electron and its $p^\downarrow$ counterpart.
This influences the overlaps of these $p^\uparrow$/$p^\downarrow$ orbitals with unoccupied Cr $e_g^\uparrow$/$e_g^\downarrow$ orbitals ($\sigma$ bonding) in the perpendicular N--Cr direction and in total effect, the FM 90\textdegree{} Cr--N--Cr superexchange arises.
Therefore, the differences between the strengths of the NN exchange parameters in AFM CrN, shown in Fig.~\ref{fig:dist}, originate in the competition between the AFM and FM contributions, each of a specific power-law dependence on the actual Cr--Cr distance.

\bibliography{main}

\end{document}